\documentclass[aps,reprint,showpacs,superscriptaddress,
footinbib,citeautoscript]{revtex4-1}

\usepackage{graphicx}

\usepackage[utf8]{inputenc}
\usepackage{csquotes}
\usepackage[american]{babel}
\usepackage[T1]{fontenc}
\usepackage{enumerate}
\usepackage{mdwlist}
\usepackage[activate=normal]{pdfcprot}
\usepackage{bbding}
\usepackage{color}
\usepackage{amsmath}
\usepackage{eqnarray,amsmath}
\usepackage{amssymb}
\usepackage{amsmath}
\usepackage{amsfonts}
\usepackage{mathrsfs}
\usepackage[titletoc,title]{appendix}
\usepackage{bm}
\usepackage{dcolumn}
\usepackage{color}
\usepackage[colorlinks=true,citecolor=blue]{hyperref}
\hypersetup{colorlinks=true,citecolor=blue,linkcolor=red
,urlcolor=blue}

\newcommand{\beq}{\begin{equation}}
\newcommand{\eeq}{\end{equation}}
\newcommand{\beqa}{\begin{eqnarray}}
\newcommand{\eeqa}{\end{eqnarray}}
\newcommand{\pr}{^{\prime}}
\begin{document}

\title{Hall viscosity for optical phonons}

\author{Shiva Heidari}
\affiliation{School of Physics, Institute for Research in Fundamental Sciences, IPM, Tehran, 19395-5531, Iran}
\author{Alberto Cortijo}
\affiliation{Instituto de Ciencia de Materiales de Madrid, CSIC, Cantoblanco, 28049  
Madrid, Spain.}
\author{ Reza Asgari}
\email{asgari@ipm.ir}
\affiliation{School of Physics, Institute for Research in Fundamental Sciences, IPM, Tehran, 19395-5531, Iran}
\affiliation{School of Nano Science, Institute for Research in Fundamental Sciences, IPM, Tehran, 19395-5531, Iran}

\begin{abstract}
We generalize the notion of dissipationless, topological Hall viscosity tensor to optical phonons in thin film Weyl semimetals. By using the strained Porphyrin thin film Weyl semimetal as a model example, we show how optical phonons can couple to Weyl electrons as chiral pseudo gauge fields. These chiral vector fields lead to a novel dissipationless two-rank viscosity tensor in the effective dynamics of optical phonons whose origin is the chiral anomaly. We also compute the contribution to this two rank Hall viscosity tensor due to the presence of an external magnetic field, whose origin is the conventional Hall response of Weyl electrons. Finally, the phonon dispersion relations of the system at the long-wavelength limit with and without an electromagnetic field are calculated showing a measurable shift in the Raman response of the system. Our results can be investigated by Raman scattering or infrared spectroscopy by attenuated total reflectance experiments.
\end{abstract}

\maketitle

\section{Introduction}\label{sec:intro}
 
One of the most interesting aspects of the physics of Weyl semimetals (WSMs) is the presence of an axionic term in their electromagnetic response, consequence of the so-called chiral anomaly~\cite{armitage2018weyl}:
\beq
\mathcal{S}=\frac{e^2}{8\pi^2}\int d^3\bm{r}dt \epsilon^{\mu\nu\rho\sigma}b_{\mu}A_{\nu}\partial_{\rho}A_{\sigma}.\label{axionaction}
\eeq
where $b_\mu=(b_0,\bm{b})$ is the constant axial four vector and ${\bm b}$ denotes the separation of two Weyl nodes of opposite chirality in momentum space. The parameter $2b_0$ represents the separation between the nodes in energy. It is customary to use the expression (\ref{axionaction}) to define the chiral anomaly $\partial_{\mu}J^{\mu}_5=\frac{e^2}{2\pi^2}\bm{B}\cdot\bm{E}$, and the expression for the anomalous Hall effect (AHE), $\bm{J}=\frac{e^2}{2\pi^2}\bm{b}\times\bm{E}$. Also one might naively derive the expression for the chiral magnetic effect (CME) in equilibrium from Eq. (\ref{axionaction}) \cite{Fukushima2008,zyuzin2012topological}. The presence of the so-called Bardeen counterterms in the effective consistent action renders the CME in equilibrium to be zero \cite{Landsteiner2016}.

The AHE derived from Eq.~(\ref{axionaction}) is a non-dissipative, parity-odd response associated with the topological structures in WSMs. But interestingly, these topological structures modify the dissipative transport and optical properties of WSMs. The paradigmatic case of these transport phenomena is the anomaly-related positive magnetoconductivity~\cite{nielsen1983adler,son2013chiral,burkov2014chiral}. The experimental observation of this effect has fueled the research activity in WSMs in the recent years. In a similar fashion, the planar Hall effect associated to the anomaly has been theoretically predicted and experimentally observed \cite{Nandy2017,Kumar2018,Li2018}. In most cases, these transport phenomena can be understood within the effective framework of chiral kinetic theory, where electrons are treated semiclassically, and the effective equations of motion are modified by the presence of the Berry curvature taking the form of a magnetic monopole in momentum space. The chirality $s=\pm1$ associated with each nodal point in WSMs can be considered as the charge of such Berry monopoles. The topological notion of WSMs can thus be understood as the (local) topological protection of such monopole charges, that can be only added or removed in pair \cite{nielsen1981absence}.
\\
It turns out that the coupling between electrons and the electromagnetic fields in WSMs is not the only that gets modified by the topological structures in the system.
Recently it has been shown that strain can couple to electrons in WSMs in the form of chiral vector fields (a coupling with
opposite sign at different nodal points). When a chiral vector field $A^{5}_{\mu}$ is taken into account, a term similar to Eq. (\ref{axionaction}) appears in the effective response of WSMs~\cite{Hutasoit2014}:
\beq
\Delta\mathcal{S}=\frac{1}{8\pi^2}\int d^3\bm{r}dt \epsilon^{\mu\nu\rho\sigma}b_{\mu}A^5_{\nu}\partial_{\rho}A^5_{\sigma},\label{axialaxionaction}
\eeq
where $b_{\mu}$ is the constant part of the axial vector field $A^5_{\mu}$. This expression has been considered before when magnetic fluctuations played the role of chiral vector fields in magnetic WSMs~\cite{Hutasoit2014}.

The presence of chiral vector fields through the action Eq.~(\ref{axialaxionaction}) not only results in a change in the expression of the chiral anomaly, $\partial_{\mu}J^{\mu}_5=\frac{e^2}{6\pi^2}(3\bm{B}\cdot\bm{E}+\bm{B}_5\cdot\bm{E}_5)$, but it also modifies the effective acoustic phonon dynamics~\cite{shapourian2015viscoelastic,cortijo2015elastic, cortijo2016visco,sinner2019spontaneous}. In the absence of time-reversal symmetry, a nonozero phonon Hall viscosity (a non-dissipative, parity-odd \emph{four-rank} tensor $\eta^{H}_{ijlr}$ appearing in the effective continuum elasticity theory)~\cite{barkeshli2012dissipationless,Hoyos2014review}. These elastic gauge fields constitute a realization of chiral magnetic fields inducing Landau levels in the absence of external magnetic fields~\cite{pikulin2016chiral, massarelli2017pseudo,Nica2018,Liu2017}.
It also contributes to the non-equilibrium chiral magnetic effect \cite{Cortijo2015strainCME}, to an unconventional mixing of acoustic phonons with plasmons in WSMs~\cite{Gorbar2017,Gorbar2017prl}, or even they induce emergent excitations in the fermionic spectrum~\cite{Wurff2019}. 

The Hall viscosity term appears in the effective, long-wavelength continuum description of elasticity, and its presence is ultimately determined by symmetries (although the precise form of the Hall viscosity coefficient might depend on the microscopic details of the system). This universality makes possible to define such effective generalized elasticity theory not only in electronic systems, but also in other realizations of Weyl systems~\cite{Ferreiros2018,Roy2018,Peri2019}.

Optical phonons, in contrast, display a behavior that strongly depends on the particular underlying lattice structure of the material, not following a universal low-energy dynamics. Nevertheless, the impact of the chiral anomaly has been already studied in some systems where the lattice structure allows for a pseudoscalar representation of phonons that couples as a chiral charge imbalance, activating the non-equilibrium CME \cite{Song2016,Rinkel2017,Rinkel2019} leading to the coupling between acoustic phonons and plasmons. 
In the present paper, we develop a theory of coupling between Weyl electrons and optical phonon modes through chiral vector fields taking a recently proposed model of Weyl fermions in a crystal of Porphyrin \cite{yuen2014topologically,owerre2016chiral}. We will compute a non-dissipative, parity odd \emph{two-rank} tensor $\eta^{H}_{ij}$ that plays the same role for optical phonons that the Hall viscosity for acoustic phonons. We explore how an external magnetic field also induces a contribution to this optical phonon Hall viscosity in a way reminiscent to the appearance of a classical Hall conductivity in the electromagnetic response of metals. While the former topological contribution is fixed, the latter contribution due to the magnetic field depends on its orientation. We compute the shift in the optical phonon frequencies for different magnetic field orientations, and discuss how these optical phonon frequency shifts can be detected by Raman spectroscopy.
It is important to note that, in the present case, the photons do not develop an effective electric dipole through the anomaly, or the induced mixing between phonons (as in Refs.\cite{Rinkel2017,Rinkel2019}) but this phonon Hall viscosity appears due to the chiral vector nature of the electron-phonon coupling.

The paper is organized as follows. In Sec.~\ref{sec:model}, we present the model Hamiltonian of an unstrained system and calculate a pseudovector potential of the system where Porphyrin thin film WSMs is deformed. The chiral charge anomaly and current densities in the presence of deformed lattice are provided in Sec. \ref{sec: chiral}. Furthermore, in Sec \ref{sec:optic phonons}, we consider the deformed system in the presence of an external magnetic and the pseudomagnetic fields, and calculate the effective action to investigate the optical phonon dispersions, numerically. We conclude and summarize our main results in Sec.~\ref{sec:conclusions}.

\section{Theory and model}\label{sec:model}

In this section we will first describe the effective electronic tight binding Hamiltonian of an undistorted three dimensional array of layers made of Porphyrin molecules~\cite{yuen2014topologically,owerre2016chiral}, and then we will consider elastic distortions, and their impact on the hopping parameters.
We will consider the Porphyrin molecules to be rigid neglecting internal molecular distortions.

It is important to notice that we decided to use the Porphyrin molecule lattice for convenience as a simple lattice system hosting Weyl points in the low-energy spectrum and simultaneously constituting a simple lattice where the optical phonon band structure can be treated analytically. In principle, the analysis developed in the following sections can be performed in any lattice system hosting Weyl fermions.

\subsection{Lattice model Hamiltonian}

\begin{figure}
\includegraphics[width=9.cm]{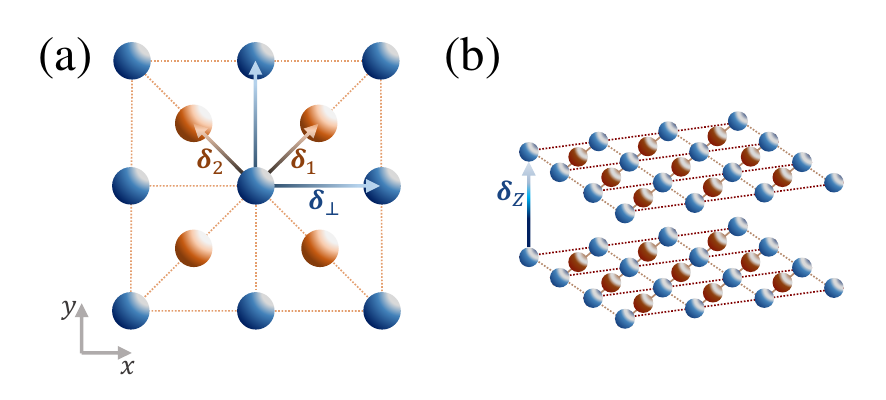}
\caption{(Color online) The Porphyrin thin film lattice from (a) top view and (b) side view. This square lattice model consists of two sublattices, namely ${a}$ (blue) sites which located at $\bm{n}_a= a(n_x,\ n_y,0)$, for $n_x, n_y$ integers, and ${b}  $ (brown) sites at $\bm{n}_b \equiv a(n_x+1/2, n_y+1/2,0)$. There are three possible hoppings between, nearest neighbor(brown arrow in (a)), next nearest neighbor(blue arrow in (a)) and between, interlayers (blue arrow in (a)).}\label{fig1}
\end{figure}

We will consider a two-dimensional bipartite square lattice of Porphyrins at positions $\bm{n}_a=a(n_x,n_y,0)$, and $\bm{n}_b=\bm{n}_a+\bm{\delta}_i$ ($i=1,2$). More details of this two dimensional model can be found in Ref. \cite{yuen2014topologically}. We then consider the effective Hilbert space composed of the lowest molecular electronic states with the corresponding creation (annihilation) operators $a_n$ ($a^{\dagger}_n$), and $b_{n}$ ($b^{\dagger}_n$) (see Fig. \ref{fig1}). The same as Haldane honeycomb model proposed the quantum Hall effect which results from a broken time reversal symmetry without any net magnetic flux \cite{haldane19883}, this model is also constituted under an inhomogeneous magnetic field with no effective magnetic flux per unit cell which reveal non-trivial topological properties. Therefore, we will consider also the presence of an external magnetic field inducing magnetic dipole transitions between the two mentioned states, that can be ultimately mapped to an effective magnetic flux per sublattice plaquette~\cite{Malley1968}, leading to complex hopping parameters $J_1$ and $J_2$, together with an effective staggered potential parametrized by $\mu_{xy}$:

\par
\begin{equation} \label{eq1}
\begin{split}
{\cal H}_{2D}=&\sum _{n}[J_1(e^{i \phi} a^\dagger_n b_{n+\hat{\delta}_1}+e^{-i \phi} a^\dagger_n b_{n-\hat{\delta}_1})\\
&+J_2(e^{-i \phi} a^\dagger_n b_{n+\hat{\delta}_2}+e^{i \phi} a^\dagger_n b_{n-\hat{\delta}_2})+h.c.]
\\
&+J_{\perp} \sum_n [a^\dagger_n a_{n+\delta_{\hat{x}(y)}}-b^\dagger_n b_{n+\delta_{\hat{x}(y)}}+h.c.]\\
&+\mu_{xy}\sum_n [a^\dagger_n a_n - b^\dagger_n b_n].
\end{split}
\end{equation}

The phase $\phi$ in Eq.(\ref{eq1}) is an effective (synthetic) flux phase that originates from the dipolar interactions between the tilted magnetic dipole operators between neighboring Porphyrin molecules (due to the time-reversal symmetry breaking effect induced by the effective magnetic flux). By itself, it is not an observable magnetic flux, so it can be modified by suitable gauge transformations. What is physical is the total circulation over closed loops. The trade of tilted magnetic dipole interactions in a homogeneous magnetic field by the description of an inhomogeneous effective magnetic field leading to effective magnetic fluxes with homogeneous hopping interactions have been considered in other systems, like dipolar spin interactions in optical lattices~\cite{Yao2012}. From now on, we set this phase to be $\phi=\pi/2$~\cite{owerre2016chiral}.

Following Ref. \cite{owerre2016chiral}, we then consider a stack of layers at the distance $d$ with a coupling phenomenologically described by a interlayer hopping Hamiltonian along the $z$ direction:
\begin{equation}
\begin{split}
{\cal H}_{inter}=&J_D \sum_n [a^\dagger_n a_{n+\delta_{\hat{z}}}-b^\dagger_n b_{n+\delta_{\hat{z}}}+h.c.]\\
&+\mu_{z}\sum_n [a^\dagger_n a_n - b^\dagger_n b_n]
\end{split}
\end{equation}
\par
The low-energy Hamiltonian of the present system is the Weyl fermion Hamiltonian \cite{owerre2016chiral} ($\hbar=1$) is given by
\begin{equation}\label{Weyleffective}
{\cal H}_0(q)={\cal H}_{2D}+{\cal H}_{inter}=v q_1 \sigma_1-v q_2 \sigma_2 + s v_3 q_3 \sigma_3
\end{equation}
where $\bm{q}=\bm{k}-\bm{b}_s $, $v=a t_{ab}$, $v_3=t_D d \sin(|\bm{b}_s| d)/2$ with the rescaled parameters $J_\perp \rightarrow -t_\perp$ and $J_D \rightarrow -t_D/4$ and $t_{ab}\rightarrow  Re {J_{1,2}}$
. The parameters $s=\pm 1$ and $\mathbf{b}_s$ denote respectively the chirality and the position in momentum space of the Weyl nodes:
\begin{equation} \label{eq10}
\bm{b}_s=(0, 0, s b)=(0, 0, \dfrac{s}{d} \cos^{-1}(\dfrac{2}{t_D}(4t_\perp -\mu)))
\end{equation}
We note that the model in Eq.~(\ref{Weyleffective}) describes an anisotropic Weyl fermion model with different velocities. This anisotropy is not essential in all the subsequent discussions, and it can be easily incorporated in the calculations by rescaling arguments.   
\subsection{Elastic gauge fields}

In order to illustrate the methodology we first review how to obtain the vector electron-acoustic phonon coupling in the system at hand. Based on the approach, we will apply the same method to obtain the vector coupling between electrons optical phonons. The strain is modeled based on the deformation of a tight-binding model. This approach  has been applied in graphene and other two-dimensional materials~\cite{guinea2010energy, vozmediano2010gauge, rostami2015theory} and in lattice realizations of Weyl semimetals~\cite{cortijo2015elastic,Cortijo2015strainCME}.

In the acoustic case, all atoms within a given unit cell move coherently and so all the atomic displacements $\bm{u}_a$ with respect to the equilibrium positions $\bm{\delta}_a$ are in the same direction, slightly changing in amplitude. We then consider that these displacements modify the hopping parameters, understood as changes in the overlap integrals of atomic orbitals at different positions.
\begin{equation}\label{acousticdisplacements}
\delta^\prime_{i,a}=\delta_{i,a}+\sum_j \epsilon_{ij} \delta_{j,a},
\end{equation}
where $\bm{\delta}_1=a(1, 1, 0)/2$, $\bm{\delta}_2=a(-1, 1, 0)/2$, $\bm{\delta}_R=a(1, 0, 0)$, $\bm{\delta}_u=a(0, 1, 0)$ and $\bm{\delta}_z=d(0, 0, 1)$. Note that in a two-dimensional system or in the case of negligible
out-of-plane deformation, $\epsilon_{ij}=\dfrac{1}{2}(\partial_i u_j + \partial_j u_i)$ is the symmetric strain tensor, where in this case we suppose all the elements are positive and nonzero.
The modification of the hopping parameters between the considered molecular orbitals can be phenomenologically parametrized by~\cite{Ishikawa2006,Ando2006}:
\begin{equation}\label{acoustichopping}
t^\prime_a=t_a(1-\beta(\dfrac{|\bm{\delta}^\prime_a|}{a}-1))
\end{equation}
where $|\bm{\delta}^\prime_a|$ are the new (deformed) relative interatomic distances. We  assume for simplicity that the Gr\"{u}neisen parameters $\beta=-\partial \log t/\partial \log \delta$ are equal for all intermolecular hopping amplitudes.
We then insert the displaced positions (\ref{acousticdisplacements}) into Eq.~(\ref{acoustichopping}) and expand in powers of the strain tensor $\epsilon_{ij}$.
Note that there is also the correction to the magnetic flux, $\phi$, owing to the area change of the unit cell after distortion and thus we have
\begin{equation}
\begin{split}
S\rightarrow &|\delta_A+\epsilon_\parallel \delta_A-\delta_B-\epsilon_\parallel \delta_B|^2=
S (1+\epsilon_\parallel)^2 \\
&=(1+2 \epsilon_\parallel +O(\epsilon^2)) S
\end{split}
\end{equation}
where $\epsilon_\parallel=\epsilon_{xx}+\epsilon_{yy}$. Consequently, the magnetic flux changes due to the deformation as
\begin{equation}
\phi \rightarrow (1+2\epsilon_{\parallel}) \phi
\end{equation}
or
\begin{equation}
J e^{i \phi} \rightarrow J e^{i \phi} e^{2i\epsilon_{\parallel} \phi}= J e^{i \phi} [1+2i \epsilon_{\parallel}\phi]
\end{equation}
Inserting these modifications into the original lattice Hamiltonian (Eq.~\ref{eq1}), the elastic gauge filed or pseudovector potential will arise. The low-energy effective Hamiltonian in the continuum limit in the vicinity of the nodal points is given by
\begin{equation} \label{eq8}
{\cal H}= v (q_x+sA^{el}_1)\sigma_x - v (q_y+sA^{el}_2)\sigma_y+s v_3 (q_z+sA^{el}_3)\sigma_z
\end{equation}
where strain couples with the low-energy electron excitations by the following vector fields
\begin{equation}
\begin{split}
&A^{el}_1= |\bm{b}_s| \epsilon_{31}+\gamma_1(\epsilon_{11} +\epsilon_{22})\\
 &A^{el}_2= |\bm{b}_s| \epsilon_{32}- \gamma_2 (\epsilon_{11}+\epsilon_{22})\\
 &A^{el}_3= |\bm{b}_s| \epsilon_{33}+\gamma_3(\epsilon_{11}+\epsilon_{22}).
\end{split}
\end{equation}
Here $\gamma_1=\tilde{\beta} / v$, $\gamma_2=(2 \pi/a+\tilde{\beta}/v)$, $\gamma_3=4 \beta/v_3$ and $\tilde{\beta}=4(\sqrt{2}-1)\pi \beta$. Note that $|\bm{b}_s|$ is the Weyl node distance in $k$ space ( given by Eq. \ref{eq10}). As expected, these vector fields chiraly couple to the electrons around the Weyl nodes.


In the case of the optical displacements, we consider relative displacements among molecules within each unit cell.
Applying the same reasoning in Eq.~(\ref{acoustichopping}) for relative displacements, we obtain~\cite{Ishikawa2006,Ando2006}:
\begin{equation} \label{eq9}
t(\bm{\delta}^\prime)=t(1+\beta\dfrac{1}{|\bm{\delta}|^2} \bm{\delta} \cdot \bm{u}(r)),
\end{equation}
where $\bm{\delta}$ and $\bm{\delta} '$ are the lattice constants in the absence and presence of strain, respectively, and $\bm{u}(\bm{r})$ is the \emph{relative} vector displacement between the two sublattices, $\bm{u}\sim \bm{u}_{a}-\bm{u}_{b}$. It means that
$\bm{\delta}'_1=\bm{\delta}_1+(u_x, u_y, u_z)$, $\bm{\delta}'_2=\bm{\delta}_2+(-u_x, u_y, -u_z)$, $\bm{\delta}'_R=\bm{\delta}_R+(u_x, u_y, -u_z)$, $\bm{\delta}'_u=\bm{\delta}_u+(-u_x, u_y, u_z)$ and finally $\bm{\delta}'_z=\bm{\delta}_z+(u_x, u_y, u_z)$.

\begin{figure}
\includegraphics[width=9cm]{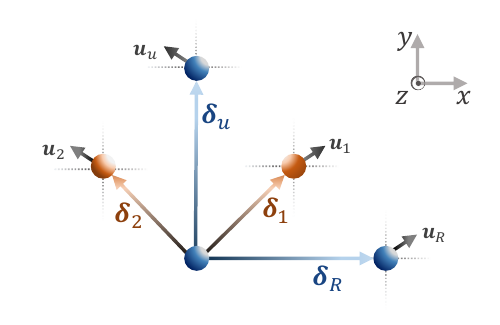}
\caption{(Color online) Arrangement of the vector displacements due to the exerted strain. The vector $\vec{u}(r)$ is the relative displacement.}\label{fig2}
\end{figure}

We consider an arrangement of the atom movements as shown in Fig. \ref{fig2}.
There is also a correction to the magnetic flux because of the area distortion of unit cell.
\begin{equation}
\phi \rightarrow \phi (1+4\dfrac{u_x}{a})=\phi (1+\epsilon_{op})
\end{equation}
or
\begin{equation}
J e^{i \phi} \rightarrow J e^{i\phi} e^{i \epsilon_{op} \phi}=J e^{i\phi} (1+i \epsilon_{op} \phi)
\end{equation}
\par
Following the same procedure as in the case of acoustic phonons, the elastic gauge fields couple to the low-energy sector of the electrons (Eq. \ref{eq8}). The corresponding elastic gauge fields are
\begin{equation}
\begin{split}
&A^{el}_1=\dfrac{2|\bm{b}_s| u_3}{a}\\ &A^{el}_2=\dfrac{4\pi u_1}{a^2}\\ &A^{el}_3=\dfrac{|\bm{b}_s|}{d}u_3-\dfrac{2}{a v} \beta(u_2+u_1).\label{opticalgauge}
\end{split}
\end{equation}

\section{Chiral charge anomaly and current densities}\label{sec: chiral}
\subsection{Determination of the anomalous response to electromagnetic and axial fields}
As we have seen, optical phonons can also couple to the electronic degrees of freedom through elastic chiral vector fields. Then, it is possible to obtain a Hall viscous response for optical phonons from the chiral anomaly as it was done for acoustic phonons. The effective action ${\mathcal{S}}_{eff}$ in terms of the electromagnetic and axial vector fields can be obtained by integrating out fermions in the adequately regulated field theory. However, there is another way to construct this effective action by analyzing the conservation laws of the currents involved in the system. This path has been discussed in the literature of Weyl semimetals in the recent years, and we quote it here for the sake of completeness.
The first step is to determine the form of the three point correlation functions that give rise to the anomalies for chiral fermions. The result is well known and given by
\beq
\partial_{\mu}J^{\mu}_s=s\frac{1}{32\pi^2}\epsilon^{\mu\nu\rho\sigma}F^{s}_{\mu\nu}F^{s}_{\rho\sigma},
\eeq
where $s=\pm1\equiv L,R$ stand for left $(+)$ and right $(-)$ chiral fermions and $F_{\mu\nu}$ is the electromagnetic tensor. We have also artificially defined left and right vector fields that couple to the corresponding chiral fermionic species by defining $J_{\mu}=\sum_s J^s_{\mu}=J^{+}_{\mu}+J^{-}_{\mu}$, and $J^5_{\mu}=\sum_s sJ^{s}_{\mu}=J^{+}_{\mu}-J^{-}_{\mu}$ as the electric and chiral currents respectively, and $A_{\mu}=A^{+}_{\mu}+A^{-}_{\mu}$, $A^5_{\mu}=A^{+}_{\mu}-A^{-}_{\mu}$, as the associated electromagnetic and chiral fields.

It has been pointed out that when both vector and axial vector fields are coupled to Weyl fermions, anomalies \emph{naively} appears in both in the chiral and in the electromagnetic sector as follows;
\begin{subequations}
\beq
\frac{\partial \rho_5}{\partial t}+\bm{\nabla}\cdot\bm{J}_5=\frac{1}{2\pi^2}\left(\bm{E}\cdot\bm{B}+\bm{E}_5\cdot\bm{B}_5\right),\label{covariant1}
\eeq
\beq
\frac{\partial \rho}{\partial t}+\bm{\nabla}\cdot\bm{J}=\frac{1}{2\pi^2}\left(\bm{E}_5\cdot\bm{B}+\bm{E}\cdot\bm{B}_5\right),\label{covariant2}
\eeq
\end{subequations}
In this precise form, the currents described in (\ref{covariant1}) and (\ref{covariant2}) represent the \emph{covariant} form of the anomalies. 
As we have an unwanted anomalous term in the electric conservation law (\ref{covariant2}), it must be fixed. The way it is proceed is by adding a so-called Bardeen polynomial to the covariant currents and it is given by
\begin{subequations}
\beq
\mathcal{J}_{\mu}=J_{\mu}-\frac{1}{4\pi^2}\epsilon^{\mu\nu\rho\sigma}A^5_{\nu}F_{\rho\sigma},\label{consistent1}
\eeq
\beq
\mathcal{J}^5_{\mu}=J^5_{\mu}-\frac{1}{12\pi^2}\epsilon^{\mu\nu\rho\sigma}A^5_{\nu}F^5_{\rho\sigma}.\label{consistent2}
\eeq
\end{subequations}

Now, the consistent currents $\mathcal{J}_{\mu}$ and $\mathcal{J}^5_{\mu}$ satisfy
\begin{subequations}
\beq
\partial_{\mu}\mathcal{J}^{\mu}=0,
\eeq
\beq
\partial_{\mu}\mathcal{J}^5_{\mu}=\frac{1}{12\pi^2}(3\bm{E}\cdot\bm{B}+\bm{E}_5\cdot\bm{B}_5).
\eeq
\end{subequations}

The current $\mathcal{J}_{\mu}$ is now conserved. As both terms accompanying $J_{\mu}$ and $J^5_{\mu}$ in Eqs.~(\ref{consistent1}, \ref{consistent2}) come from the same Bardeen polynomial $\Delta\mathcal{S}_{eff}$ in the effective action, we now consider both $\mathcal{J}_{\mu}$ and $\mathcal{J}^5_{\mu}$ as the \emph{physical} currents in our problem.

Now, the important observation is that the \emph{covariant} currents $J_{\mu}$ and $J^5_{\mu}$ are the currents obtained by using chiral kinetic theory, when a finite Fermi level $\mu$ is taken into account at zero temperature, so we will compute these currents using this formalism. Before that, it is important to make a last remark. In the right hand side of Eq. (\ref{consistent2}), the term originating from the Bardeen polynomial appears to be quadratic in the axial vector field $A^5_{\nu}$, however, we have to remember that, formally, the positions of the Weyl nodes enter as a part of this axial field: $\bm{A}^5_{\mu}=b_{\mu}+A^{el}_{\mu}$. Then, we can rewrite Eq. (\ref{consistent2}) as
\beq
\mathcal{J}^5_{\mu}\simeq J^5_{\mu}-\frac{1}{12\pi^2}\epsilon^{\mu\nu\rho\sigma}b_{\nu}F^{el}_{\rho\sigma}+...,\label{consistent3}
\eeq
after keeping the linear order in the phonon vector field $A^{el}_{\mu}$. We are interested in the current $J^5_{\mu}$ which is the current that will enter in the effective action of optical phonons, so we will devote ourselves to its computation in the next paragraphs.

Before moving into the next discussion, we would like to state that $\phi$ appeared in Eq. (3) should be modified, in principle, when the system is addressed by an external magnetic filed owing  to the Zeeman term. The Porphyrin lattice unit cell contains two molecules. The lowest energy Hilbert subspace of each molecule comprises three states: the ground-state and two degenerate $Q-$ bands with the spin singlet state, ${\bf S}=0$. Therefore, the angular momentum is the only term that contributes to the Zeeman term in the low-energy effective Hamiltonian. The Zeeman splitting is proportional to the cosine of the angle between the magnetic field and the axis perpendicular to the molecule plane~\cite{yuen2014topologically}. Although, the Zeeman term is negligible when the applied magnetic field is parallel to the Porphyrin surface, it has a very small contribution when the magnetic field is applied along the surface~\cite{Malley1968}. Putting numbers, if $|{\bf B}|=1$ mT, the ratio of the Zeeman splitting to the energy level spacing of the Porphyrin is around $10^{-6}$. Therefore, we ignore the Zeeman coupling in the effective Hamiltonian.

\subsection{Computation of the consistent currents using chiral kinetic theory}
As  mentioned above, we consider our Weyl semimetal at finite chemical potential. For sufficiently large Fermi energies, we can neglect inter-band transitions and make use of the chiral kinetic effective description of the response of electrons to external perturbations.

We will compute the averaged chiral current $\bm{J}^5=\sum_{s}s\bm{J}_s$ as a response to the phonon chiral vector field $\bm{A}^{5}=\sum_{s}s\bm{A}^{el}$. Then, writing the response of each chiral fermionic species to the axial vector field in terms of a polarization tensor, we can write $J^5_i=\sum_s s^2 \Pi_{ij}^s A^{el}_j$, with $s^2=1$, so only the part of $\Pi^s_{ij}$ that does not depend on the chirality index $s$ will survive to the summation. This part is the conventional one that can be found in several places in the literature.

We will solve the Boltzmann equation in the collisionless limit $\omega\tau\gg1$, where $\tau$ is the lifetime of quesiparticles, to ease the discussion. As usual, we will compute the probability density in the phase space per chiral specie $f_s$:
\beq
\dot{f}_s+\dot{\bm{x}}_s\cdot\partial_{\bm{x}}f_s+\dot{\bm{k}}_s\cdot\partial_{\bm{k}}f_s=0,\label{boltzmann1}
\eeq

with the equations of motion 
\begin{subequations}
\beq
D_s \dot{\bm{x}}_s=\bm{v}^s-es^2\bm{\Omega}\times\bm{E}^{el}+s e(\bm{v}^s\cdot\bm{\Omega})\bm{B},\label{EOMx}
\eeq
\beq
D_s \dot{\bm{k}}_s=s\bm{E}^{el}+e\bm{v}^s\times\bm{B}+s^2e(\bm{E}^{el}\cdot\bm{B})\bm{\Omega},\label{EOMk}
\eeq
\end{subequations}
where we have used $\bm{E}^{el}=\dot{\bm{A}}^{el}$, $\bm{\Omega}^s=s\bm{\Omega}$ for our Weyl metal model, $D_s=1+s e\bm{B}\cdot\bm{\Omega}$ as the volume modification of the phase space, and we have neglected the contribution of the pseudomagnetic field $\bm{B}^{el}=\bm{\nabla}\times\bm{A}^{el}$. We are allowed to utilize it as the optical phonons which are gapped modes and we are interested in the modification of this gap, working then in the local limit, $\bm{A}^{el}(\omega)\sim \bm{A}^{el}e^{i\omega t}$. With this simplification, and multiplying Eq. (\ref{boltzmann1}) by $D_s$ we thus have

\beq
i\omega D_s f_s+D_s\dot{\bm{k}}_s\cdot\partial_{\bm{k}}f_s=0,\label{boltzmann2}
\eeq

As usual, we consider the linear response regime where corrections to the equilibrium distribution function $f_0(\varepsilon_{\bm{k}})$ that are linear in the $\bm{A}^{el}$ field are taken into account, $f_s=f_0+\frac{\partial f_0}{\partial \varepsilon}\delta f_s$:
\beqa
i\omega D_s \delta f_s&+&(e\bm{v}^s\times\bm{B})\cdot\partial_{\bm{k}}\delta f_s=-(s\bm{v}^{0}\cdot\bm{E}^{el}+\nonumber\\
&+&e\bm{v}^{0}\cdot\bm{\Omega}(\bm{E}^{el}\cdot\bm{B})).\label{boltzmann3}
\eeqa
\subsection{External magnetic field}

In the presence of an external magnetic field, the dispersion relation gets modified by the presence of the orbital magnetic moment $\bm{m}^s_{\bm{k}}$, namely $\varepsilon_{\bm{k}}=\varepsilon^0_{\bm{k}}+e\bm{B}\cdot\bm{m}^s_{\bm{k}}$ where $\bm{m}_k^s=-\epsilon_k \bm{\Omega}_k^s=-s v \hat{\bm{k}} / 2|k|$ comes from the self-rotation of the Bloch wave packet around its center. The corresponding group velocity is
\beq
\bm{v}_s=\vec{\nabla} \epsilon_k=v \hat{\bm{k}} (1+2 es (\bm{B} \cdot \bm{\Omega}))-e s v \bm{B} (\hat{\bm{k}} \cdot \bm{\Omega})
\eeq
with $\hat{\bm{k}}=\bm{k}/k$ and assuming $v_3=v_1=v_2=v$. Also, for the isotropic Weyl model, we have $\bm{\Omega}=\frac{1}{2}\frac{\hat{\bm{k}}}{k^2}$. We will assume low temperature approximation where we can use the Sommerfeld expansion. As the chiral current is
\beq
\bm{J}^5=\frac{1}{8\pi^3}\sum_s s\int d^3\bm{k} D_s \dot{\bm{x}}_s \frac{\partial f_0}{\partial \varepsilon}\delta f_s,\label{current1}
\eeq
We will restricted to the small deviation from Fermi energy due to the low temperature. 
Thus, we can write $D_s=1+s e\bm{B}\cdot\bm{\Omega}=1+s eB\frac{1}{2}\frac{|\hat{k}_3|}{k^2}=1+s\alpha \cos\theta$, after defining the dimensionless parameter $\alpha=e B v^2/2\mu^2$. In the same circumstances, $\bm{v}^{0}\cdot\bm{\Omega}=v^3/2\mu^2$, and the differential operator $\hat{\Theta}=(e\bm{v}^s\times\bm{B})\cdot\partial_{\bm{k}}$ simply reads $\hat{\Theta}=eB \frac{v^2}{\mu} (1+2 s \alpha \cos \theta)\partial_{\varphi}\equiv\omega_c(1+2 s \alpha \cos \theta) \partial_{\varphi}$. The cyclotron frequency is $\omega_c=eB v^2/\mu$.\par 

With all these simplifications, we can write the Boltzmann equation (\ref{boltzmann3}) in the following form
\beqa
&&i\omega(1+s\alpha \cos\theta) \delta f_s+\omega_c (1+2 s \alpha \cos \theta) \partial_{\varphi}\delta f_s=\nonumber\\
&-&(sv\bm{E}^{el}\cdot\hat{\bm{k}}+v\alpha E^{el}_3).\label{boltzmann4}
\eeqa
We make use of the standard parametrization of vectors in spherical coordinates. We then decompose $\delta f_s$ in harmonics as $\delta f_s=\delta f^0_s+\delta f^+_s e^{i\varphi}+\delta f^-_s e^{-i\varphi}$ and obtain the following solutions.
We make use of the standard parametrization of vectors in spherical coordinates. We then decompose $\delta f_s$ in harmonics as $\delta f_s=\delta f^0_s+\delta f^+_s e^{i\varphi}+\delta f^-_s e^{-i\varphi}$ and obtain the following solutions.
\begin{subequations}
\beq
\delta f^0_s=i v\frac{(s\cos\theta+\alpha)}{\omega(1+s \alpha \cos\theta)} E^{el}_3,
\eeq
\beq
\delta f^+_s= is v\frac{\sin\theta}{\omega(1+s \alpha \cos\theta)+\omega_c (1+2 s \alpha \cos \theta)} E^{*}_{el},
\eeq
\beq
\delta f^-_s=i s v\frac{\sin\theta}{\omega(1+s \alpha \cos\theta)-\omega_c (1+2 s \alpha \cos \theta)} E_{el},
\eeq
\end{subequations}
with $E_{el}=\frac{1}{2}(E^{el}_1+i E^{el}_2)$, and $E^*_{el}=\frac{1}{2}(E^{el}_1-i E^{el}_2)$.

The chiral current in Eq. (\ref{current1}) can be further simplified in the same terms as the Boltzmann equation (\ref{boltzmann4})
\beq
J^{5}_i=-\frac{\mu^2+\frac{\pi^2}{3}(k_B T)^2}{8\pi^3 v^2}\sum_s s\int^{2\pi}_0 d\varphi \int^{1}_{-1} du \hat{k}_i (1+2s\alpha u)\delta f_s,\label{current2}
\eeq
with the change of variables $u=\cos\theta$.

Splitting the current (\ref{current2}) for each harmonics, we have, for the $0$th harmonic

\beqa
J^{5(0)}_i && =-i\frac{ \mu^2+\frac{\pi^2}{3}(k_B T)^2}{4\pi^2 v \omega}  \\ \nonumber & \times & \sum_s\int^1_{-1}du \frac{u(1+2s \alpha u)(u+s \alpha)}{(1+s\alpha u)}\delta_{i3}E^{el}_3.\label{zerothcurrent1}
\eeqa

In order to have a finite expression in Eq. (\ref{zerothcurrent1}), $\alpha$ might be between zero and unity. 
As we are interested in the regime of small magnetic fields, described by the hierarchy of scales $l_F\ll l_B$ (magnetic length much larger that the Fermi wavelength), we have $\alpha\ll 1$ so we expand the integrand in Eq. (\ref{zerothcurrent1}) in powers of $\alpha$. Integrating and summing over $s$, we have
\beqa
\bm{J}^{5}_{\parallel}=\frac{  \mu^2+\frac{\pi^2}{3}(k_B T)^2}{3\pi^2 v} \left(1+\frac{2}{5}\alpha^2\right)\bm{A}^{el}_{\parallel}.\label{zerothcurrent2b}
\eeqa
$\bm{J}^{5}_{\parallel}$ and $\bm{A}^{el}_{\parallel}$ stand for the parallel components to the magnetic field.
From the previous expression we can read off the parallel component of the chiral polarization tensor
\beqa
\Pi_3=\frac{\mu^2+\frac{\pi^2}{3}(k_B T)^2}{3\pi^2 v} \left(1+\frac{2}{5}\alpha^2\right).\label{zerothcurrent2c}
\eeqa
The contribution proportional to $\alpha^2\sim B^2$ is the same contribution to the positive magnetoconductivity in Weyl semimetals. This term is always positive, and, being associated to the longitudinal part of the chiral polarization tensor, will renormalize the optical phonon gap.

We can proceed in the same way with the other harmonic components, obtaining, after integration upon $\varphi$:
\begin{subequations}
\beqa
J^{5(+)}_i&=&\frac{-i }{8\pi^2 v} (\mu^2+\frac{\pi^2}{3}(k_B T)^2) \left(\begin{array}{c}
1 \\
i \\
0\end{array}\right) E^{*}_{el}\nonumber\\
&\times&\sum_s\int^{1}_{-1}du \frac{(1+2s \alpha u)(1-u^2)}{\omega(1+\alpha s u)+\omega_c (1+2 s \alpha u)},\label{1stcurrent2}
\eeqa
\beqa
J^{5(-)}_i&=&\frac{-i }{8\pi^2 v} (\mu^2+\frac{\pi^2}{3}(k_B T)^2)\left(\begin{array}{c}
1 \\
-i \\
0\end{array}\right) E_{el}\nonumber\\
&\times&\sum_s\int^{1}_{-1}du \frac{(1+2s \alpha u)(1-u^2)}{\omega(1+\alpha s u)-\omega_c (1+2 s \alpha u)}\label{2ndcurrent2}
\eeqa
\end{subequations}

As before, we expand these expressions up to second order in $\alpha$ and sum over chiralities. After integrating upon $u$ and collecting terms,  we finally have

\beqa
\bm{J}^5_{\perp}=\Pi_0\bm{A}^{el}_{\perp}+\Pi_{H}\bm{B}\times \bm{A}^{el}_{\perp},
\eeqa
with 

\beqa
\Pi_0= &&\frac{2\omega^2 (\mu^2 +\frac{\pi^2}{3} (k_B T)^2)}{3\pi^2 v} \cdot\\ \nonumber && \left(\frac{1}{\omega^2-\omega^2_c}- \frac{\alpha^2\omega^2}{5}\frac{\omega^2-3\omega^2_c}{(\omega^2-\omega^2_c)^3}\right),
\label{pi0}
\eeqa

\beqa 
\Pi_H= && \frac{-2eiv\omega (\mu^2 +\frac{\pi^2}{3} (k_B T)^2)}{3\pi^2 \mu} \cdot \\  && \nonumber \left(\frac{1}{\omega^2-\omega^2_c}-\frac{\alpha^2\omega^2}{5}\frac{\omega^2-5\omega^2_c}{(\omega^2-\omega^2_c)^3}\right).\label{pih}
\eeqa 

$\bm{J}^5_{\perp}$ and $\bm{A}^{el}_{\perp}$ denote the part of the current and vector fields that are perpendicular to $\bm{B}$, respectively.

The final form of the chiral polarization tensor is
\beq
\Pi_{ij}=\Pi_3\delta_{i3}\delta_{j3}+\Pi_0\delta^{\perp}_{ij}+\Pi_H\varepsilon_{lij}B_l.\label{chiralpolarization1}
\eeq

From Eq. (\ref{consistent3}) we can read the temperature independent \emph{anomaly-related} contribution to the chiral polarization tensor as
\beq
\Delta \Pi_{ij}=-\frac{i\omega}{12\pi^2}\varepsilon^{3ij}|\bm{b}|.\label{chiralpolarization2}
\eeq

\begin{figure}
	\includegraphics[width=8 cm]{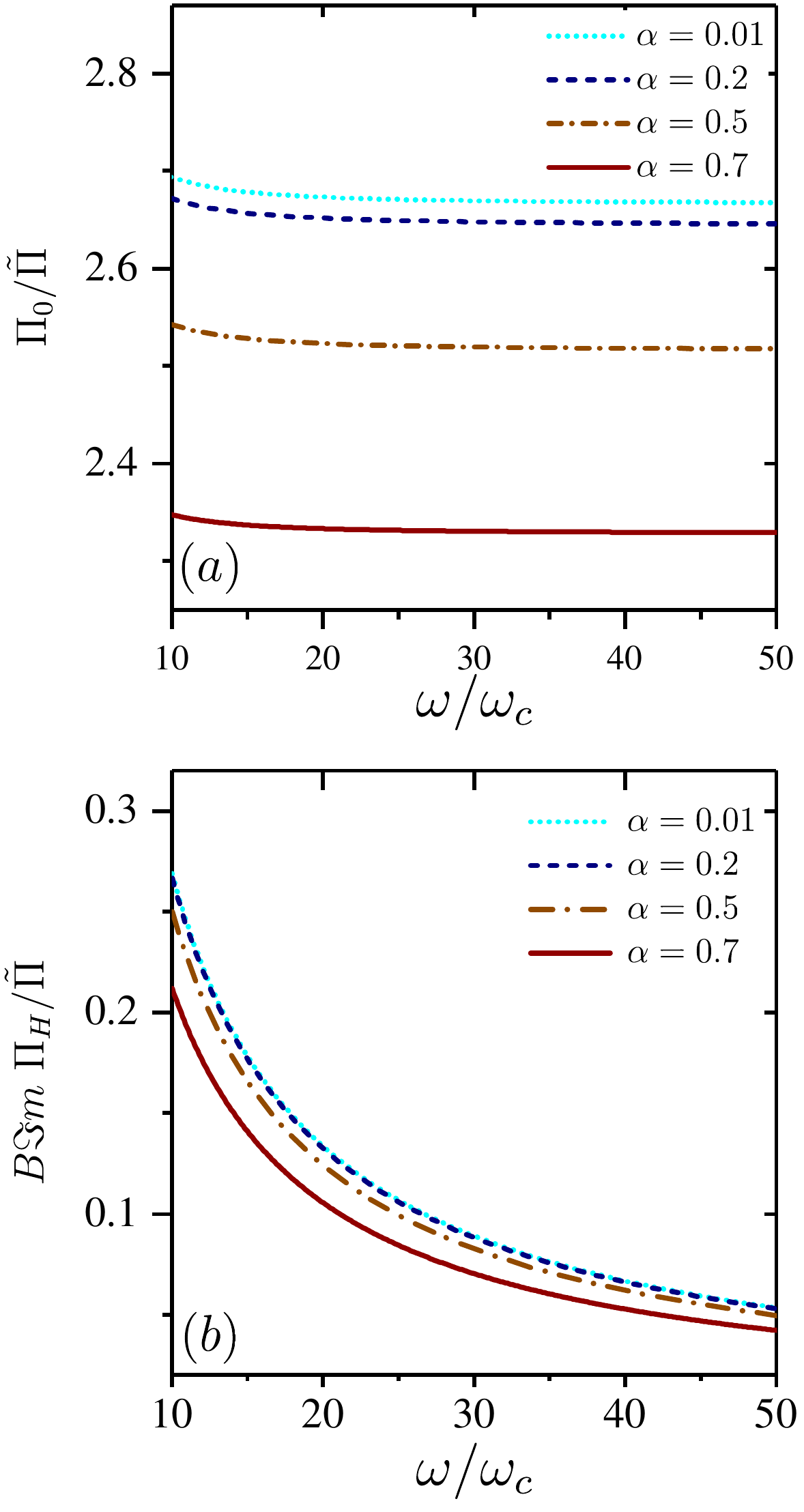}
	\caption{(Color online) Numerical results of (a) $\Pi_0$ and (b) $\Im m\Pi_H$ as a function of the frequency (in units of $\omega_c$) for different values of $\alpha=eB v^2 /2\mu^2$ with $\tilde{\Pi}=(\mu^2+\frac{\pi^2}{3} (k_B T)^2)/{8 \pi^2 v}$. Notice that $\Pi_0$ is almost constant in terms of the frequency, however $\Im m\Pi_H$ disperse strongly. Those parameters decrease by increasing the external magnetic field. }\label{fig3}
\end{figure}

    Here, we would like to explore $\Pi_0$ and $\Pi_H$ given by Eqs. (\ref{1stcurrent2}) and (\ref{2ndcurrent2}) as a function of the frequency for different values of $\alpha$. As shown in Fig.~\ref{fig3}, $\Pi_0$ changes very slightly when increasing the frequency, however $\Im m\Pi_H$ decreases fast. Those parameters increase with growing the external magnetic field. It would be worth mentioning that our numerical results fully cover analytical expressions given by Eqs. (37) and (38) for a very small $\alpha$ value. 

\section{Dynamics of the longitudinal optical phonon mode}\label{sec:optic phonons}

The previous expressions for the polarization tensor in Eqs. (\ref{chiralpolarization1}, \ref{chiralpolarization2}) constitute the (local) response of Weyl electrons to the chiral vector field made of \emph{difference} of displacements of the two sub-lattices (that eventually will lead to the optical phonon displacements (\ref{opticalgauge})). In terms of the displacement components, this vector can be written in a matrix form as 
\beqa
A^{el}_{i}=\Lambda^{a}_{i}u_a\equiv\left(\begin{array}{ccc}
0 & 0 &\frac{2b}{a} \\\
\frac{4\pi}{a^2} & 0 &0 \\\
-\frac{2\beta a}{v} & -\frac{2\beta a}{v} & \frac{b}{d}\end{array}\right)\left(\begin{array}{c}
u^{a}_1-u^{b}_1\\
u^{a}_2-u^{b}_2\\
u^{a}_3-u^{b}_3\end{array}\right).
\eeqa

This chiral polarization can be seen as a piece of the effective action for phonons:

\beqa\nonumber
\Delta \mathcal{L}&=&A^{el}_{i}(\Pi_{ij}+\Delta \Pi_{ij})A^{el}_{j}=\\ &=& \nonumber
(\Pi_3\hat{B}_i\hat{B}_j+  \Pi_0\delta^{\perp}_{ij}+\Pi_H  \varepsilon_{lij}B_l-\frac{i\omega}{12\pi^2}\varepsilon^{3ij}|\bm{b}|) \\
&\times&   \Lambda^{r}_i\Lambda^{s}_{j} (u^{a}_r-u^{b}_r)(u^{a}_s-u^{b}_s),\label{totaleffectivelagrangean}
\eeqa

\begin{figure}
\includegraphics[width=7.5 cm]{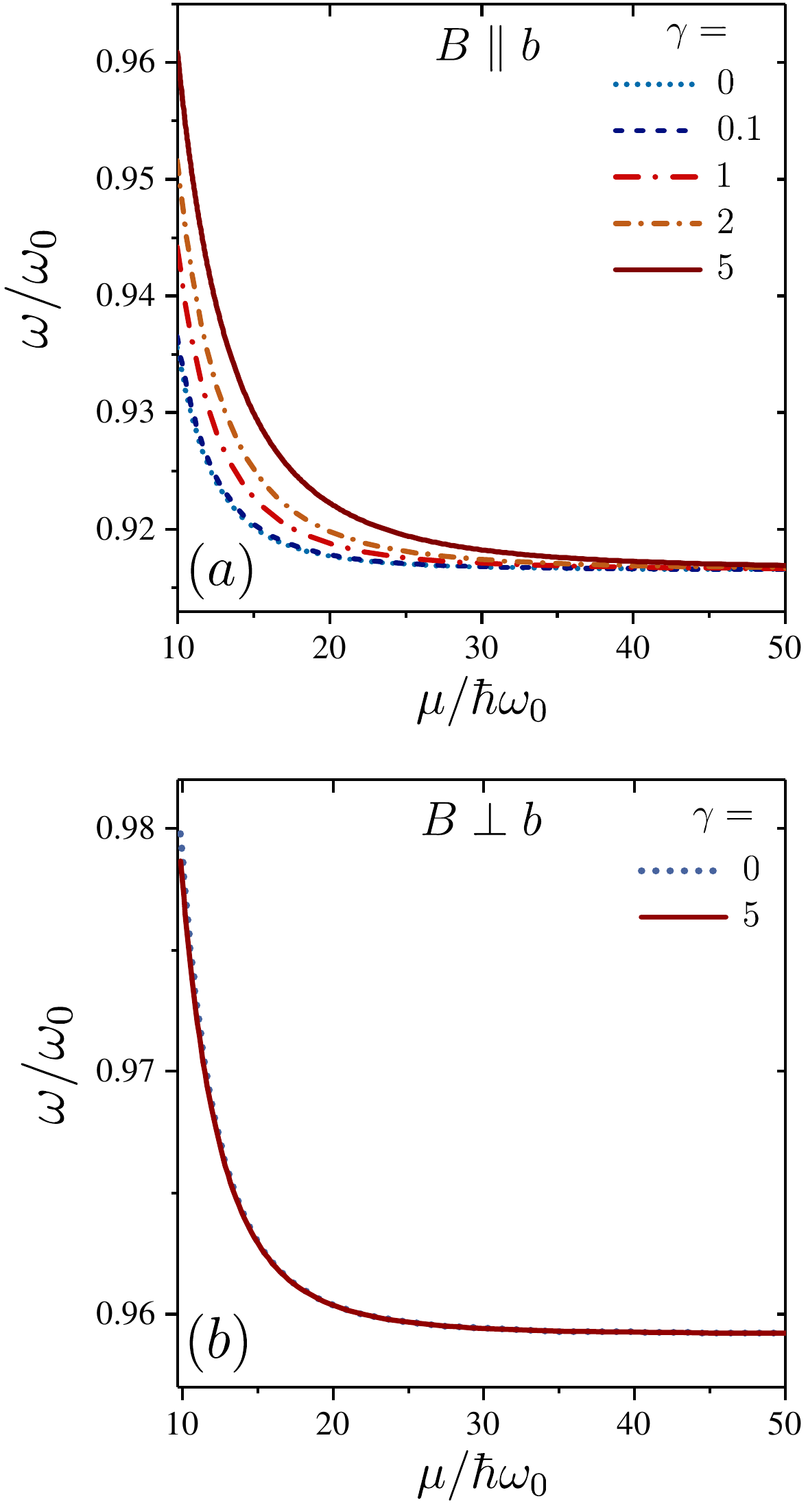}
\caption{(Color online) Normalized frequency of the optical phonon at k=0 as a function of the  chemical potential (in units of $\hbar\omega _c$) for different values of $\gamma=eB v^2/\hbar \omega_0^2$ ($\omega_0$ being the finite optical frequency of a centered square lattice at k=0 see Appendix. \ref{appa}). $\gamma=0$ denotes a zero magnetic field. We set $d=2 a$, $a^2 \beta / t_{ab}=1$, $b=(5 d)^{-1}$, $v=10^6$ m/s, $\hbar\omega_0=6.5$ meV and $M_a=M_b$.(a) When magnetic field, $\bm{B}$, is in a parallel to the Weyl node separation vector, $\bm{b}$, the phonon frequency rises by the Fermi energy decrement (but still interband transitions are negligible) and shift upward by $\gamma$ increment.(b) The shift of the phonon frequency in the case that $\bm{B} \perp \bm{b}$ is vanishingly small. }\label{fig4}
\end{figure}

\begin{figure}
	\includegraphics[width=7.5 cm]{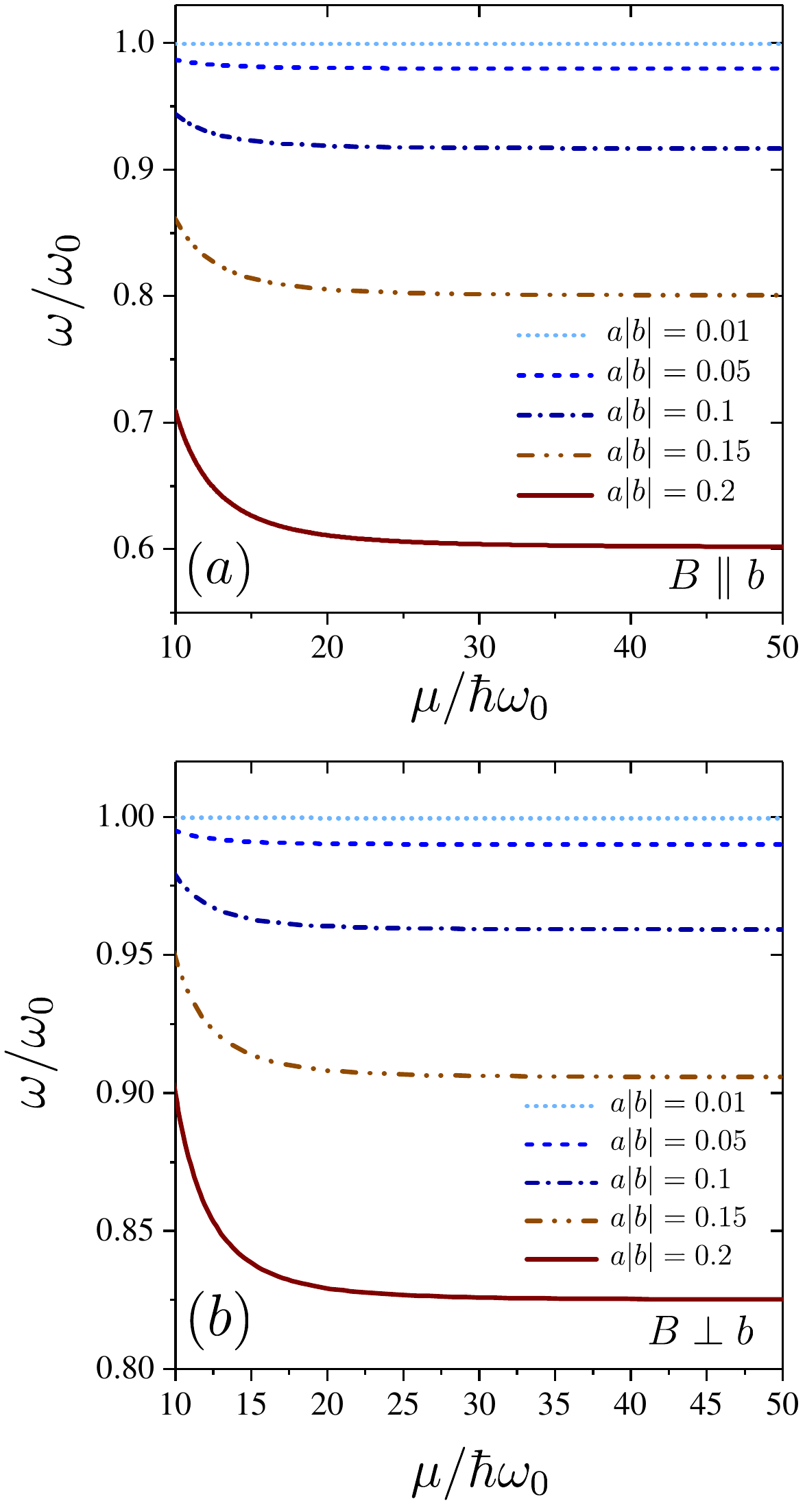}
	\caption{(Color online) Normalized frequency of optical phonons at $k=0$ for different values of $a |b| $. $a |b|$ is coefficient of some gauge fields terms, so distinct values of $a |b|$ are corresponding to changing the strain intension.  We set $d=2 a$, $a^2 \beta / t_{ab}=1$, $M_a=M_b$, $\gamma=1$, $\hbar\omega_0=6.5$ meV and $v=10^6$ m/s.}\label{fig5}
\end{figure}

where $\hat{B}_i$ is the unit vector pointing along the magnetic field $\bm{B}$. The symbol $\delta^{\perp}_{ij}$ refers to a Kronecker delta only in the indices \emph{perpendicular to the magnetic field}. Also, it is important to remember that we are considering the local approximation, where electrons react to the presence of the optical deformations \emph{within} each unit cell. 

We are interested in the change of the optical phonon frequencies at $\bm{k}=0$ (at the $\Gamma$ point) due to the Hall-like part of the phonon action in Eq. (\ref{fig5}). These shifts can be measured in Raman experiments (we present the details of the calculations in Appendix A). 

Similar studies have been performed for acoustic phonons~\cite{barkeshli2012dissipationless,cortijo2016visco}. Here we will follow the same strategy: The anomaly-induced polarization in Eq.~(\ref{chiralpolarization2}) is insensitive to the magnetic field and fixed in the direction of $\bm{b}$, in contrast to Eq.~(\ref{chiralpolarization1}). We can apply the external magnetic field in different directions, and see the change in the optical phonon frequencies. We first choose the magnetic field to point parallel to the direction of the separation of the Weyl points $\bm{b}$ ($\bm{B}=B \hat{\bm{z}}$) and then we choose $\bm{B}$ such that $\bm{B} \perp \bm{b}$.

The Fermi energy dependence of the optical phonon at $\bm{k}=0$ is shown in Fig.~\ref{fig4} for different values of $\gamma=eBv^2/\hbar \omega^2_0$ where $\omega_0$ is the bare optical phonon frequency in the case that the magnetic field is (a) parallel or (b) perpendicular to the Weyl node axis. Notice that the kinetic approach is valid only in the regime where $\omega_c \tau \ll 1$ and $\hbar \omega_c \ll \epsilon_{\rm F}$ where $\tau$ is the lifetime of the quasiparticles. The optical phonon decreases with increasing the electron density and it changes in the magnetic field when $\bm{B} \| \bm{b}$.  
\\
It is essential to mention that $\Pi_0$ and $\Pi_3$ have a dominate contribution on the coupled phonon frequency shift at the $\Gamma$ point and Hall viscose term $\Pi_H$ causes the frequency $\omega$ to move up $\omega_0$. However, the results show that  $\Delta\Pi_{ij}/k^2_{\rm F}$ makes the coupled optical phonon disperses against the Fermi energy.   

In addition, we show that by reducing $b$ values (corresponding to reducing the value of $\Delta\Pi_{ij}$ explicitly), the optical phonon becomes flatter and attains to unity as illustrated in Fig.~\ref{fig5}. Another physical interpretation of $b$ is the fact that it denotes the strength of strain. Larger $b$ means a larger strain value. For a given magnetic field, the $\omega/\omega_0$ decreases by increasing $b$.

\section{Conclusions}\label{sec:conclusions}

In the present paper we have developed the theory of anomaly-induced Hall viscosity for optical phonons, by constructing an explicit example where optical phonons couple to electrons through an elastic chiral vector field. Then, a dissipationless, parity-odd Hall viscosity appears in the phonons dynamics. This dynamic is inherited from the topological nontrivial response to vector fields of the low energy electronic semi metallic band structure, that can be described by the Weyl Hamiltonian. The Hall viscosity for optical phonons is then yet another probe to test the topological Berry curvature in Weyl systems.

The mechanism presented in the present work is different from other mechanisms of Hall viscosity generation published in the literature~\cite{Rinkel2017,Song2016,Rinkel2019}. In these works, an explicit use of the chiral anomaly is used, but several terms in the electron-phonon coupling (vector, chiral vector, and pseudoscalar) are required to trigger the anomaly and obtain a closed equation for the optical phonon dynamics. Such requirements severely constraint the crystalline groups (and thus the materials) where that realization can be observed. In the present case, by contrast, only the absence of time reversal symmetry is required to trigger the version of the anomaly comprised of $\bm{b}$ and two chiral vector fields $\bm{A}^{el}_5$~\cite{Hutasoit2014}.

The experimental observability Hall viscosity of acoustic phonons appears to be challenging~\cite{barkeshli2012dissipationless,cortijo2015elastic,Liu2017}.
We have found that the presence of an external magnetic field induces an extra term to the Hall viscosity through a more conventional mechanism, similar to the Hall conductivity induced by the Fermi surface in metals. The presence of this term allows us to use the external magnetic field as a tuning parameter to measure the optical phonon Hall viscosity. While the conventional Hall contribution strongly depends on the particular direction of the magnetic field $\bm{B}$, the contribution from the anomaly does not, as shown in Fig.~\ref{fig4}. This proposes a method to measure the optical phonon frequency shift for different magnetic field orientations to extract the anomaly contribution of the Hall viscosity. 

We again stress that the results presented in this paper are not particular on the system analyzed, that was chosen for analytical tractability. These results can be generalized to any material hosting Weyl fermions by studying the corresponding changes in the band structure due to the optical lattice deformations. Similar analysis can be carried out in other magnetic Weyl semimetals, as Mn$_3$Sn \cite{Kuroda2017}, Co$_3$Sn$_{2}$S$_2$ \cite{Liu2018}, or Co$_2$MnGa \cite{Sakai2018}.

It is interesting to observe that our results are in conceptual agreement with the analysis performed in two spatial dimensions, where a Chern-Simons term for the optical vector fields $\bm{A}^{el}$ is dynamically induced by massive Dirac electrons~\cite{Sinner2016}.

Finally, as a technical comment, it is worth to mention that, for simplicity, we used an isotropic low-energy model for Weyl fermions (Eq. (\ref{eq8})), in a manifestly anisotropic system. Considering an anisotropic Weyl semimetal does not change our qualitative conclusions, although the qualitative values for the Hall viscosity for optical phonons might vary. It is easy to incorporate velocity anisotropies in our calculations by appropriate rescaling the momentum integrals. This translates into overall geometric factors depending only of the velocities in front of each component of the polarization tensor. Then, one needs to go to the effective phonon Lagrangean (Eq. \ref{totaleffectivelagrangean}) and compute again the phonon frequency shifts.

\section{acknowledgments}
We thank M. Vozmediano and K. Landsteiner for very useful discussions. This work is supported by the Iran Science Elites Federation. A.C. acknowledges financial support through MINECO/AEI/FEDER, UE Grant No. FIS2015-73454-JIN and European Union structural funds and the Comunidad Autonoma de Madrid 
(CAM) NMAT2D-CM Program (S2018-NMT-4511).

\newpage

\begin{widetext}
	\appendix
	\section{Phonon modes in a force constant model (No viscous terms)} \label{appa}
	We start with the approach of the force constant model. In this approach we consider spring instead of interatomic forces.
	Let us consider our lattice model is a stacking layers with a square shape consists of Prophyrins located into two sublattices A and B in the absence of EM field and ignoring Hall viscosity properties in the system.
	\\ In general, the equation of motion of the displacement of $a^{th}$ atom ($a=A,B$), placed at  the site $\bm{d}_a$ within the unit cell labeled by $\bm{R}_n$ is given by:
	\begin{equation}
	M_a \ddot{\bm{u}}^{a}(\bm{R}_n+\bm{d}_a)= \sum_{b}\sum_{n\pr} K^{(ab)} (\bm{u}^{b}(\bm{R}_{n}+\bm{R}_{n\pr}+\bm{d}_b)-\bm{u}^{a}(\bm{R}_{n}+\bm{d}_{a})).\label{dynamicalequation1}
	\end{equation}
	In our particular lattice, we will set $\bm{d}_1=\bm{0}$, and $\bm{d}_2=\bm{\delta}_1$. Also, each position has four nearest neighbor sites belonging to the other sublattice in the plane, and two nearest neighbors of the same sublattice in the third direction. We will denote $K^{(ab)}=\kappa_1$ when considering interactions among sites of different sublattices, $a\neq b$, and $K^{(ab)}=\kappa_2$ for interactions among sites belonging the same sublattice.
	
	We will project Eqs. (\ref{dynamicalequation1}) along the corresponding lattice vectors involved in each term, and after Fourier transforming $(\bm{u}^{a}= e^{i\bm{k}\bm{R}_n-i\omega t}\bm{u}^{a}_{\bm{k}})$:
	\beqa
	-M_a \omega^2 \bm{u}^{a}=\kappa_1\sum_{j}e^{i\bm{k}\cdot \hat{h}_j}(\bm{u}^{b}\cdot\hat{h}_j)\hat{h}_{j}-\kappa_1\sum_{j}(\bm{u}^{a}\cdot\hat{h}_j)\hat{h}_{j}+
	\kappa_2\sum_{j\pr}\left(e^{i\bm{k}\cdot\bm{n}_{j\pr}}-1\right)(\bm{u}^{a}\cdot\hat{n}_{j\pr})\hat{n}_{j\pr}.
	\eeqa
	It is useful to write this equation in components ($\hat{n}_{j\pr}$ is the orthogonal basis set):
	\beqa
	\omega^2 M_a u^{a}_s+\kappa_2\sum_{j\pr}\left(e^{i\bm{k}\cdot\bm{n}_{j\pr}}-1\right)u^{a}_s-\kappa_1\left(\sum_{j} \hat{h}^{j}_{s} \hat{h}^{j}_r\right)u^{a}_{r}+\kappa_1\left(\sum_j e^{i\bm{k}\cdot\bm{h}_j}\hat{h}^{j}_{s}\hat{h}^{j}_r\right)u^{b}_r=0.\label{phononequation1}
	\eeqa
	In this notation, we have rewritten $\bm{n}_1=\bm{\delta}_u=-\bm{n}_3$, $\bm{n}_2=\bm{\delta}_R=-\bm{n}_4$, $\bm{n}_3=\bm{\delta}_z=-\bm{n}_6$, $\bm{h}_1=\bm{\delta}_1=-\bm{h}_3$, and $\bm{h}_2=\bm{\delta}_2=-\bm{h}_4$.
	
	To see how the phonon frequency at the $\Gamma$ point gets affected by the induced viscosities, it is enough to set $\bm{k}=0$ in Eq. (\ref{phononequation1}):
	\beqa
	\omega^2 M_a u^{a}_s-\kappa_1\left(\sum_{j}\hat{h}^{j}_{s}\hat{h}^{j}_r\right)u^{a}_{r}+\kappa_1\left(\sum_j \hat{h}^{j}_{s}\hat{h}^{j}_r\right)u^{b}_r=0.\label{phononequation2}
	\eeqa
	
	Using the expressions for $\hat{h}^j$, we simply have $\sum_{j}\hat{h}^{j}_{s}\hat{h}^{j}_r= \delta_{rs}$ so, for the in-plane displacements, we have:
	
	\beqa
	(\omega^2 M_a-\kappa_1)u^{a}_{s}+\kappa_1 u^{b}_s=0.\label{phononequation3}
	\eeqa
	
	The solution of this equation gives the standard text-book values for the optical phonon frequencies:
	\beq
	\omega_{TO,LO}(\bm{k}=0)=\sqrt{\kappa_1}\sqrt{\left(\frac{1}{M_a}+\frac{1}{M_b}\right)}.\label{Optfreq}
	\eeq


\section{Phonon equation of motion in the presence of chiral vector fields}
In the case $\bm{B} \parallel \bm{b}$ ($\vec{B}= B \hat{z}$), the polarization tensor as a response of the phonon chiral vector field would be
	\begin{equation}
	\Pi_{ij}+\Delta \Pi_{ij}=
	\begin{pmatrix}
	\Pi_0 && B \Pi_H + \Delta && 0 \\
	-B \Pi_H - \Delta && \Pi_0 && 0 \\
	0 && 0 && \Pi_3
	\end{pmatrix}
	\end{equation}
	with $\Delta=-i \dfrac{\omega}{12 \pi^2} b$. Phonon equations of motion after adding the terms coming from the Lagrangian is given by 
	\begin{equation} \label{a2}
	\begin{split}
	& (\omega^2 -\kappa_1 ) u^a_1+\kappa_1 u^b_1+\Pi_0 \Lambda^1_2 \Lambda^1_2  (u_1^a-u_1^b)+\Pi_3 \Lambda_3^1 \Lambda_3^1 (u^a_1-u^b_1)+\Pi_3 \Lambda_3^1 \Lambda_3^2 (u^a_2-u^b_2)+\\ &+ \Pi_3 \Lambda^1_3 \Lambda^3_3 (u^a_3-u^b_3)+(B \Pi_H+\Delta \Pi) \Lambda^3_1 \Lambda^1_2 (u^a_3-u^b_3)=0 \\
	\\
	& (\omega^2 -\kappa_1 ) u^b_1+\kappa_1 u^a_1-\Pi_0 \Lambda^1_2 \Lambda^1_2 (u_1^a-u_1^b)-\Pi_3 \Lambda_3^1 \Lambda_3^1 (u^a_1-u^b_1)-\Pi_3 \Lambda_3^1 \Lambda_3^2 (u^a_2-u^b_2)-\\ & -\Pi_3 \Lambda_3^3 \Lambda_3^1 (u^a_3-u^b_3)-(B \Pi_H+\Delta \Pi) \Lambda_1^3 \Lambda^1_2 (u^a_3-u^b_3)=0 \\
	\\
	& (\omega^2 -\kappa_1 ) u^a_2+\kappa_1 u^b_2+\Pi_3 \Lambda^2_3 \Lambda^1_3 (u_1^a-u_1^b)+\Pi_3 \Lambda_3^2 \Lambda_3^2 (u^a_2-u^b_2)+\Pi_3 \Lambda_3^2 \Lambda_3^3 (u^a_3-u^b_3)=0\\ 
	\\
	& (\omega^2  -\kappa_1 ) u^b_2+\kappa_1 u^a_2-\Pi_3 \Lambda^1_3 \Lambda^2_3 (u_1^a-u_1^b)-\Pi_3 \Lambda_3^2 \Lambda_3^2 (u^a_2-u^b_2)-\Pi_3 \Lambda_3^2 \Lambda_3^3 (u^a_3-u^b_3)=0 \\
	\\
	& (\omega^2  -\kappa_1 ) u^a_3+\kappa_1 u^b_3+\Pi_0 \Lambda^3_1 \Lambda^3_1 (u_3^a-u_3^b)+\Pi_3 \Lambda_3^3 \Lambda_3^1 (u^a_1-u^b_1)+\Pi_3 \Lambda_3^3 \Lambda_3^2 (u^a_2-u^b_2)+\\ & +\Pi_3 \Lambda_3^3 \Lambda_3^3 (u^a_3-u^b_3)+(B \Pi_H+\Delta \Pi) \Lambda_1^3 \Lambda^1_2 (u^b_1-u^a_1)=0 \\
	\\
	& (\omega^2 -\kappa_1) u^b_3+\kappa_1 u^a_3-\Pi_0 \Lambda^3_1 \Lambda^3_1 (u_3^a-u_3^b)-\Pi_3 \Lambda_3^3 \Lambda_3^1 (u^a_1-u^b_1)-\Pi_3 \Lambda_3^2 \Lambda_3^3 (u^a_2-u^b_2)-\\ & -\Pi_3 \Lambda_3^3 \Lambda_3^3 (u^a_3-u^b_3)-(B \Pi_H+\Delta \Pi) \Lambda_1^3 \Lambda^1_2 (u^b_1-u^a_1)=0 \\
	\\
	\end{split}
	\end{equation}
where $\kappa_1=(1+M_a/M_b)^{-1}$, $\Lambda_1^3=2 b a$, $\Lambda_1^2=4 \pi$, $\Lambda_1^3=\Lambda_2^3=-2 \beta a/\nu$ and $\Lambda_3^3=a^2 b/d$ are the dimensionless quantity, and the elements of the polarization tensor are normalized to $(\mu^2+\frac{\pi^2}{3} (k_B T)^2)/3 \pi^2 v$ , and $\omega \rightarrow \omega / \omega_0$ where $\omega_0$ is the phonon mode in the absence of the viscous term. 

 The above equations is equvalent to solve this matrix equation
	\begin{equation}
	\hat{M}_{(6 \times 6)} \begin{pmatrix}
	u_1^{a,b} \\
	u_2^{a,b} \\
	u_3^{a,b}
	\end{pmatrix}_{(6 \times 1)}=0
	\end{equation}
As we are interested in a small magnetic field limit, the cyclotron frequency is not that large in compared to the finite optical phonon mode. 
We can thus expand the expressions of $\Pi_0$ and $\Pi_3$ up to the second order of $\omega_c$, and the results lead to
	\begin{equation}
	\begin{split}
	& \Pi_0=2 [(1-\dfrac{\alpha^2}{5})+(\dfrac{\omega_c}{\omega})^2] \\
	& B \Pi_H=-2 i (1-\dfrac{\alpha^2}{5} )(\dfrac{\omega_c}{\omega})
	\end{split}
	\end{equation}
The shift of the optical phonon frequency at the $\Gamma$ point is determined numerically by the requirement that the determinant of matrix $\hat{M}$ vanishes.\\
In the case $\bm{B} \perp \bm{b}$  ($\vec{B}=B \hat{x}$), we expect different shift of the phonon frequency due to the anisotropy induced by anisotopic axial gauge fields. The polarization tensor in this case is written as
	\begin{equation}
	\Pi_{ij}+\Delta \Pi_{ij}=
	\begin{pmatrix}
	\Pi_1 &&  \Delta && 0 \\
	-\Delta && \Pi_0 && B \Pi_H \\
	0 && -B \Pi_H && \Pi_0
	\end{pmatrix}
	\end{equation}
with $\Pi_1=\dfrac{\mu^2+\frac{\pi^2}{3}(k_B T)^2}{3 \pi^2 v}(1+\dfrac{2}{5} \alpha^2)$. Consequently, the phonon equations of motion when $\bm{B} \perp \bm{b}$ are different from what appeared in Eq. \ref{a2}.
	\begin{equation}
	\begin{split}
	& (\omega^2  -\kappa_1 ) u^a_1+\kappa_1 u^b_1+ \Pi_0 (\Lambda^1_2 \Lambda^1_2 +\Lambda^1_3 \Lambda^1_3) (u_1^a-u_1^b)+(\Pi_0 \Lambda^1_3 \Lambda^3_3+\Delta \Lambda^3_1 \Lambda^1_2-B \Pi_H\Lambda^1_2 \Lambda^3_3 ) (u^a_3-u^b_3)+\\ &+(\Pi_0 \Lambda^1_3 \Lambda^2_3-B \Pi_H \Lambda^1_2 \Lambda^2_3)(u^a_2-u^b_2)=0 \\
	\\
	& (\omega^2  -\kappa_1 ) u^b_1+\kappa_1 u^a_1-\Pi_0 (\Lambda^1_2 \Lambda^1_2 +\Lambda^1_3 \Lambda^1_3) (u_1^a-u_1^b)-(\Pi_0 \Lambda^1_3 \Lambda^3_3+\Delta \Lambda^3_1 \Lambda^1_2-B \Pi_H\Lambda^1_2 \Lambda^3_3 ) (u^a_3-u^b_3)-\\ &-(\Pi_0 \Lambda^1_3 \Lambda^2_3-B \Pi_H \Lambda^1_2 \Lambda^2_3)(u^a_2-u^b_2)=0 \\
	\\
	& (\omega^2  -\kappa_1 ) u^a_2+\kappa_1 u^b_2+ \Pi_0 (\Lambda^2_3 \Lambda^1_3 +B \Pi_H \Lambda^1_2 \Lambda^2_3) (u_1^a-u_1^b)+\Pi_0 \Lambda^2_3 \Lambda^2_3 (u^a_2-u^b_2)+\Pi_0 \Lambda^2_3 \Lambda^3_3 (u^a_3-u^b_3)=0 \\
	\\
	& (\omega^2 -\kappa_1 ) u^b_2+\kappa_1 u^a_2- \Pi_0 (\Lambda^2_3 \Lambda^1_3 -B \Pi_H \Lambda^1_2 \Lambda^2_3) (u_1^a-u_1^b)+\Pi_0 \Lambda^2_3 \Lambda^2_3 (u^a_2-u^b_2)-\Pi_0 \Lambda^2_3 \Lambda^3_3 (u^a_3-u^b_3)=0 \\
	\\
	& (\omega^2  -\kappa_1 ) u^a_3+\kappa_1 u^b_3+(\Pi_0 \Lambda^3_3 \Lambda^1_3 -\Delta \Lambda^3_1 \Lambda^1_2+B \Pi_H \Lambda^3_3 \Lambda^1_2) (u_1^a-u_1^b)+\Pi_0 \Lambda^2_3 \Lambda^3_3 (u^a_2-u^b_2)+ \\ & +(\Pi_0 \Lambda^3_3 \Lambda^3_3+\Pi_1 \Lambda^3_1 \Lambda^3_1)(u^a_3-u^b_3)=0 \\
	\\
	& (\omega^2 -\kappa_1 ) u^b_3+\kappa_1 u^a_3-(\Pi_0 \Lambda^3_3 \Lambda^1_3 -\Delta \Lambda^3_1 \Lambda^1_2+B \Pi_H \Lambda^3_3 \Lambda^1_2) (u_1^a-u_1^b)-\Pi_0 \Lambda^2_3 \Lambda^3_3 (u^a_2-u^b_2)- \\& -(\Pi_0 \Lambda^3_3 \Lambda^3_3+\Pi_1 \Lambda^3_1 \Lambda^3_1)(u^a_3-u^b_3)=0 \\
	\\
	\end{split}
	\end{equation}
	The phonon frequency shift at $\bm{k}=0$ is determined by solving the above matrix equation  numerically.
\end{widetext}

\bibliography{ref.bib}

\end{document}